# THE INFLUENCE OF CRACKS IN ROTATING SHAFTS

J-J. Sinou[1*] and A.W. Lees[2]

Department of Mechanical Engineering, University of Wales Swansea,
Singleton Park, Swansea SA2 8PP, Wales, UK.

[2] E-mail: a.w.lees@swansea.ac.uk
[1] E-mail: jean-jacques.sinou@ec-lyon.fr
Permanent address: Ecole Centrale de Lyon, Laboratoire de Tribologie et Dynamique des Systèmes UMR CNRS 5513, 36 avenue Guy de Collongue, 69134 Ecully, France.

## ABSTRACT

In this paper, the influence of transverse cracks in a rotating shaft is analyzed. The paper addresses the two distinct issues of the changes in modal properties and the influence of crack breathing on dynamic response during operation. Moreover, the evolution of the orbit of a cracked rotor near half of the first resonance frequency is investigated. The results provide a possible basis for an on-line monitoring system.

In order to conduct this study, the dynamic response of a rotor with a breathing crack is evaluated by using the Alternate Frequency/Time Domain approach. It is shown that this method evaluates the non-linear behaviour of the rotor system rapidly and efficiently by modeling the breathing crack with a truncated Fourier series. The dynamic response obtained by applying this method is compared with that evaluated through numerical integration. The resulting orbit during transient operation is presented and some distinguishing features of a cracked rotor are examined.

## 1. INTRODUCTION

The influence of a transverse crack in the shaft of rotating machines on the associated dynamic behaviour, has been a focus of attention for many researchers [1-35]. The presence of a crack may lead to a dangerous and catastrophic effect on the dynamic behaviour of rotating structures and cause serious damage to rotating machinery. Therefore, the timely detection of a rotor crack would potentially avoid severe damage and expensive repairs due to the failure of rotating machinery as well as assuring the safety of personnel.

Generally, two different approaches are attempted to identify the presence of a crack in rotating structures. The first approach is based on the fact that the presence of a crack in rotating shaft reduces the stiffness of the structure, hence reducing the natural frequencies of the original uncracked shaft.

Various theoretical and experimental works [3-10] performed over the last three decades have indicated that the change in modal properties (natural frequencies and modes shapes) may be useful for the detection of a crack, as well as for the identification of both crack depth and location. Moreover, the influence of opening and closing of crack due to the shaft self-weight for various orientations of the shaft has been investigated [3,11] and showed the effectiveness of the change in natural frequencies versus the orientation of the shaft to detect the orientation of the crack front. Another approach of crack identification is based on the modification of the dynamic responses of the crack rotor during its rotation. Indeed dynamic analysis of the cracked rotor based on theoretical and experimental studies has been a subject of great interest for the last decades [12-22]. Wauer [1] reviewed a literature survey on the state-of-art of the dynamics of cracked rotors. Mayes and Davies [11] analyzed the response of a multi-rotor-bearing system containing a transverse crack in a rotor both experimentally and theoretically. Gasch [12,13] studied the dynamic behaviour of a simple rotor with a cross-sectional crack and the associated stability behaviour due to the crack and imbalance. Henry and Okah-Avae [23] investigated the effects of the gravity and unbalance on the dynamic behavior of a crack shaft. Moreover, many studies [3, 20, 22, 24-28] indicated the change in dynamic responses and more particularly the tendency of the rotor to exhibit a harmonic component at twice shaft speed close to half any resonance frequency. Indeed, due to the shaft self-weight and the rotation of the rotor, the crack opens and closes during a complete revolution of the rotor; hence the stiffness of the shaft varies. This opening and closing mechanism, which is called the breathing effect, induces vibration of the second and higher harmonics of rotating speed in frequency domain. Although the presence of this twice per revolution component can indicate the possibility of the presence of a crack, it is also well known that the presence of such a component can be generated by shaft misalignment, asymmetric shaft, looseness of bolts and nuts, or a range of other non-linearities [26-27]. Therefore, this feature, in itself, is insufficient to indicate the presence of a transverse crack in rotors. In this way, the additional observation of the orbits can be useful for revealing the presence of a crack [24-29]. Effectively, near the speed range of half any resonant frequency, the orbit changes from a single loop to double loops with speed, and then an internal loop may appear. This observation is the signature of the presence of a crack that indicates the change in amplitude and phase at half any resonance speed, and is also a characteristic for signals containing two vibration components with the same direction of precession [24,27].

The aim of this paper is to investigate these effects by taking into account the non-linear dynamical behaviour due to the breathing transverse crack in order to obtain some indications that might be useful in detecting the presence of a crack in rotating system. Some parametric studies regarding the location and depth of cracks are carried out in order to show their influence on the change in frequencies. Next the influence of a transverse breathing crack on the response of a rotor model, is investigated using the Alternate Frequency/Time Domain method [36-38] with a path following procedure in order to predict the non-linear dynamics and also the twice per revolution component close to half resonance response associated with the presence of a breathing crack. It is shown that the non-linear behavior of the rotor system with a breathing crack may be obtained by modeling the crack with a truncated Fourier series. In order to validate this approach, the results are compared with those obtained through numerical integration. Moreover, it is demonstrated that a rotor with a breathing crack, which opens and closes during rotations, shows a non-linear dynamic behavior due to the variation of the rotor's stiffness during its rotation. Finally, the non-linear behavior of the crack rotor at half the first shaft resonance speed is calculated, and the influence of the crack depth is presented to provide information for an on-line identification of rotor cracks.

## 2. MATHEMATICAL MODEL

This section describes the derivation of the system model which is used in the subsequent analysis.

### 2.1. ROTOR SYSTEM

The shaft is discretized into a number of Timoshenko beam finite elements having with four degrees of freedoms at each node [39]:

$$\left(\mathbf{M}_T^e + \mathbf{M}_R^e\right)\ddot{\mathbf{x}}_i^e + \left(\mathbf{C}_R^e - \Omega\mathbf{G}_R^e\right)\dot{\mathbf{x}}_i^e + \mathbf{K}_B^e\mathbf{x}_i^e = \mathbf{Q}_i^e \tag{1}$$

where $\mathbf{M}_T^e$ and $\mathbf{M}_R^e$ are the rotary and translational mass matrix of the shaft element. $\mathbf{C}_R^e$, $\mathbf{G}_R^e$, and $\mathbf{K}_B^e$ are the external damping, gyroscopic, and stiffness matrices respectively. $\mathbf{Q}_i^e$ defines the gravity force vector for the shaft. $\Omega$ is the rotational speed and the factor of damping for the shaft. The damping is taken as classical for the sake of simplicity and $\mathbf{C}_R^e = \beta\mathbf{K}_B^e$ where the $\beta$ is a constant factor of proportionality and internal rotor damping has been neglected.
The modeling of the rigid discs is given by

$$\left(\mathbf{M}_T^d + \mathbf{M}_R^d\right)\ddot{\mathbf{x}}_i^d - \Omega\mathbf{G}^d\dot{\mathbf{x}}_i^d = \mathbf{Q}_i^d + \mathbf{W}_i^d \tag{2}$$

where $\mathbf{M}_T^d$, $\mathbf{M}_R^d$, and $\mathbf{G}_R^d$ are the translational mass, rotary mass and gyroscopic matrices respectively. $\mathbf{Q}_i^d$ consists of the weight of the disc. $\mathbf{W}_i^d$ defines the unbalance forces due to disc having mass $m$ with an eccentricity $e$.
Finally, the discrete bearing stiffness coefficients are placed at the corresponding degrees of freedom and the equation of motion for the complete rotor system is defined as follows

$$\mathbf{M}\ddot{\mathbf{x}} + \left(\mathbf{C} + \Omega\mathbf{G}\right)\dot{\mathbf{x}} + \mathbf{K}\mathbf{x} = \mathbf{Q} + \mathbf{W} \tag{3}$$

where $\mathbf{M}$ and $\mathbf{G}$ are the mass and gyroscopic matrices including mass and gyroscopic matrices of the shaft and rigid discs. $\mathbf{C}$ and $\mathbf{K}$ are the external damping and stiffness matrices of the shaft. $\mathbf{Q}$ and $\mathbf{W}$ define the vector of gravity force and imbalance force for the complete rotor system.

### 2.2. THE CRACKED ROTOR

In this section, the modeling of the crack and the breathing mechanism are discussed briefly. Mayes and Davies [3] demonstrated that a transverse crack in a rotor shaft can be represented by the reduction of the second moment of area $\Delta I$ of the element at the location of the crack. By using Rayleigh's method, they obtained that the change in $\Delta I$ verified [3,5]

$$\frac{\Delta I/I_0}{1-\Delta I/I_0} = \frac{R}{l}\left(1-\nu^2\right)F\left(\mu\right) \tag{4}$$

where $I_0$, $R$, $l$, $\nu$, $\mu$, and $F\left(\mu\right)$ are the second moment of area, the shaft radius, the length of the section, the Poisson's ratio, the non-dimensional crack depth, and the compliance functions varied with the non-dimensional crack depth $\mu$, respectively. The non-dimensional crack depth $\mu$ is given by $\mu = a/R$ where $a$ defines the crack depth of the shaft as illustrated in the Figure 1. Although $F\left(\mu\right)$ can be derived from the appropriate stress factor, Mayes and Davies indicated that a good approximation of this function can be achieved by considering the fractional change in the second moment of area measured at the crack face. Their observations were validated by experimental tests consisting of measuring the two functions $\mathrm{F}_X\left(\mu\right)$ and $\mathrm{F}_Y\left(\mu\right)$ [3] for both the $X$ and $Y$ directions indicated the direction of the crack front and the orthogonal direction associated as illustrated in Figure 1. The values of the second moment of area with the new centroid associated are given in Appendix A. Then the stiffness matrix due to the transversal crack $\mathbf{K}_{\text{crack}}$ can be obtained at the crack location, by using standard finite elements [5].

When the rotor is cracked, the opening and closing behavior due to the rotor rotation and shaft self-weight results in a time dependent stiffness. To accurately predict the dynamic response of the rotor system with a breathing crack, an appropriate crack model is essential. Many researchers have studied this problem [11-13, 17] and various crack opening and closing models have been developed. Penny and Friswell [16] compared different crack models: the hinge model (13) where the breathing switches from it open to closed state abruptly during the rotation of the shaft; the crack model of Mayes and Davies [11] where the opening and closing of the crack was described by a cosine function; and finally the crack model of Jun et all [22] based on fracture mechanics where the coupling stiffness as well as the direct stiffness are calculated as the shaft rotates (and so the crack opens and closes). Penny and Friswell [16] demonstrated that for monitoring using low frequency vibration, simple models of crack are adequate and sufficient for the prediction of the dynamic behavior of a rotor with breathing crack, as well as the predicted whirl orbit at the steady-state $2\times$harmonic of rotor speed. The extent of crack opening will be determined by the proportion of the crack face which is subject to tensile axial stresses. By assuming that the gravity force is much greater that the imbalance force, the function describing the breathing crack [1,3-4,11-13,16-17] may be chosen as

$$f(t) = \frac{1}{2}\left(1 - \cos(\Omega t)\right) \tag{5}$$

where $\Omega$ is the rotational speed of the rotor. As illustrated in the Figure 1, for $f(t) = 0$, the crack is totally closed and the cracked rotor stiffness is equal to the uncracked rotor stiffness. For $f(t) = 1$, the crack is full open.

Finally, the dynamics equation of the rotor with a breathing crack, in this linearised approximation, can be defined as

$$\mathbf{M}\ddot{\mathbf{x}} + \left(\mathbf{C} + \Omega\mathbf{G}\right)\dot{\mathbf{x}} + \left(\mathbf{K} - f(t).\mathbf{K}_{\text{crack}}\right)\mathbf{x} = \mathbf{Q} + \mathbf{W} \tag{6}$$

where $\ddot{\mathbf{x}}$, $\dot{\mathbf{x}}$, and $\mathbf{x}$ are the acceleration, velocity and displacement of the degree-of freedom of the cracked rotor system. $\mathbf{M}$, $\mathbf{C}$, $\mathbf{G}$ and $\mathbf{K}$ define the mass, damping, gyroscopic and stiffness matrices, respectively. $\mathbf{Q}$ and $\mathbf{W}$ are the vector of gravity force and imbalance force, respectively. $\mathbf{K}_{\text{crack}}$ is the stiffness matrix due to the crack and $f(t)$ the function representing the breathing effect.

In this study, a rotor shaft of 1m length and 10mm diameter; two discs of 40mm diameter and 15mm thickness are situated at 0.3m of each end of the shaft, as illustrated in Figure 2. A crack is added at one third of the left end and a mass *m* at the eccentricity *e* is placed on the first disc. All the values of the parameters are given in Table 1.

## 3. NATURAL FEQUENCIES

In this section, the changes in the natural frequencies of the rotor, which is the common first step in the diagnosis of a crack, have been examined. The crack position, depth and the orientation of the crack have been varied.

Tables 2 gives the values of the natural frequency for the uncracked and cracked shaft with the variation of the non-dimensional crack depth $\mu$. Table 3 gives the values of the natural frequency for the uncracked and cracked shaft with the variation of the crack location (with $\mu = 1$).

The natural frequencies associated with the vertical (first, third and fifth modes) and horizontal (second, fourth, and sixth modes) frequencies are equal in the case of an uncracked shaft but are different for the cracked rotor due to the presence of the crack. The highest changes in natural frequencies occur in the vertical mode due to the orientation of the crack and the shaft self-height.
In order to compare the frequencies of the cracked and uncracked shaft, the percentage change in natural frequencies is defined as follow

$$\%\mathrm{C}_i = \frac{f_i^{\text{uncracked}} - f_i^{\text{cracked}}}{f_i^{\text{uncracked}}} \times 100 \tag{7}$$

where $i$ defines the $i^{\text{th}}$-frequency of the system.
Figure 3 shows the percentages changes in the first and second natural frequencies with the rotation of the shaft for various non-dimensional crack depths $\mu$. In this case the area of the open crack due to the shaft's self-weight is calculated at each orientation of the shaft [5]. As observed previously by Lees and Friswell [5], this simple procedure is sufficient to reflect the change in natural frequencies due to the rotation of the shaft and the physics associated. If the non-dimensional crack depth $\mu$ is equal 1(corresponding to the lost of half the shaft's area), it appears that the crack is fully open at a orientation of 0 degree, and fully closed at 180 degrees, as illustrated in Figure 3(a). Moreover, one observes that for a non-dimensional crack depth $\mu$ is less than 1, the maximum change in frequencies is defined for ranges of shaft orientations that increases when the non-dimensional crack depth $\mu$ decreases. This reflects the fact that the crack and the equivalent area of the cracked shaft due to the shaft self-weight are completely open and closed for ranges of shaft orientations. Of course the change in the natural frequencies increase when the crack depth increases.
The crack depth, the location of the crack, and the rotation of the shaft clearly effects the natural frequencies of the rotor and these changes in modal properties can be used as an identification of transverse crack in rotating machines. However this approach is time consuming and requires the knowledge of the original natural frequencies of the rotor in its uncracked condition. This last point may not be easy to obtain due to the fact that since the modal tests have been carried out, the natural frequencies may have been modified during time rotating machines operation. Therefore the use of the dynamic response in order to identify the presence of a transverse crack appears to be unavoidable and the most convenient.

## 4. NON-LINEAR DYNAMICAL BEHAVIOUR

Due to the presence of a breathing crack and the mathematical model associated, the determination of the dynamic behaviour of the rotor requires considerable computational resources by using a classical numerical integration. In order to avoid these computational problems, one of the most efficient and systematic approaches is the use of the harmonic balance method [38] that permit the discretisation of the unknowns functions in time by using their Fourier components, which are assumed to be constant with respect to time. Such methods include the incremental harmonic balance method [41], the Fast Galerkin method [40,42], and the alternate frequency/time domain method [36-38].
In this study, the Alternate Frequency/Time domain method (AFT method) is applied in order to obtain the non-linear response of the rotor with the breathing crack.

### 4.1. ALTERNATE FREQUENCY/TIME DOMAIN METHOD

The system defined in equation (6) can be written more generally as

$$\mathbf{M}\ddot{\mathbf{x}} + \left(\mathbf{C} + \Omega\mathbf{G}\right)\dot{\mathbf{x}} + \left(\mathbf{K} - \mathrm{f}(t,\mathbf{x}).\mathbf{K}_{\text{crack}}\right)\mathbf{x} = \mathbf{Q} + \mathbf{W} \tag{8}$$

where there is no restriction to the unbalance magnitude.

In the linearised crack model, it is usual to represent the portion of the crack below the neutral axis as being open. There are a number of approaches differing only in detail but one such approach is that of Mayes and Davies [3]. The area of the crack face below the neutral axis is calculated and this is then represented with a chordal equivalent crack having the same area. The important approximation here is that the neutral axis is usually taken as the centroid of the shaft, but this not exact.

With any non-zero deflection, the centre line of the shaft will be stretched and hence there will be a tensile stress and this being so, for a shaft deflected under gravity, the true neutral axis will be above the centre line. For small vibration amplitudes, the error introduced by this approximation is relatively minor, simply reducing the effective crack by a constant factor. However, for appreciable vibration amplitudes the error becomes significant and time dependent. It is beyond the scope of the present paper to analyze the sensitivity of this effect but the correction (non-linear) terms are important when the vibration amplitude is appreciable relative to the catenary.

In the present paper the position of the neutral axis at the crack location has been re-calculated at each time step in order to give an accurate model of the crack dynamics.

This non-linear equation may be re-written

$$\mathbf{M}\ddot{\mathbf{x}} + \mathbf{D}(\Omega)\dot{\mathbf{x}} + \mathbf{K}\mathbf{x} + \mathbf{f}(\mathbf{x},\Omega,t) - \mathbf{g}(\mathbf{x},\Omega,t) = 0 \tag{9}$$

where $\mathbf{M}$, $\mathbf{D}$ and $\mathbf{K}$ are the mass, the damping and the stiffness matrices. $\mathbf{f}$ and $\mathbf{g}$ are the vectors containing the non-linear expressions due to the breathing crack and the vector for the imbalance and gravity force, respectively. In the absence of response to unbalance, the function $\mathbf{f}$ becomes a linear function of $\mathbf{x}$; for non-negligible imbalance, the addition stresses interact with those arising from the (gravity induced) catenary, so changing the neutral axis, and hence the active cross section of the crack.

Setting $\mathbf{x} = \mathbf{x}_k + \mathbf{\Delta x}$, $\dot{\mathbf{x}} = \dot{\mathbf{x}}_k + \mathbf{\Delta}\dot{\mathbf{x}}$ and $\ddot{\mathbf{x}} = \ddot{\mathbf{x}}_k + \mathbf{\Delta}\ddot{\mathbf{x}}$ (where $k$ defined the $k^{\text{th}}$ iteration process), one considers the truncated Fourier series expansion

$$\mathbf{x} = \mathbf{X}_0 + \sum_{i=1}^{m}\left[\mathbf{X}_{2i-1}\cos(\mathrm{i}\Omega t) + \mathbf{X}_{2i}\sin(\mathrm{i}\Omega t)\right] \tag{10}$$

where $\mathbf{X}_0$, $\mathbf{X}_{2i-1}$ and $\mathbf{X}_{2i}$ are the Fourier coefficients of $\mathbf{x}$, and the associated truncated Fourier series expansion for $\Delta\mathbf{x}$

$$\Delta\mathbf{x} = \Delta\mathbf{X}_0 + \sum_{i=1}^{m}\left[\Delta\mathbf{X}_{2i-1}\cos(\mathrm{i}\Omega t) + \Delta\mathbf{X}_{2i}\sin(\mathrm{i}\Omega t)\right] \tag{11}$$

where $\Delta\mathbf{X}_0$, $\Delta\mathbf{X}_{2i-1}$ and $\Delta\mathbf{X}_{2i}$ are the Fourier coefficients of $\Delta\mathbf{x}$. The matrix of Fourier coefficients of $\mathbf{x}$ and $\Delta\mathbf{x}$ are indicated and arranged as follows:

$$\mathbf{X} = \begin{bmatrix} X_{1,0} & \cdots & X_{i,0} & \cdots & X_{n,0} \\ \vdots & & \vdots & & \vdots \\ X_{1,2j} & \cdots & X_{i,2j} & \cdots & X_{n,2j} \\ \vdots & & \vdots & & \vdots \\ X_{1,2m} & \cdots & X_{i,2m} & \cdots & X_{n,2m} \end{bmatrix} \tag{12}$$

and

$$\Delta\mathbf{X} = \begin{bmatrix} \Delta X_{1,0} & \cdots & \Delta X_{i,0} & \cdots & \Delta X_{n,0} \\ \vdots & & \vdots & & \vdots \\ \Delta X_{1,2j} & \cdots & \Delta X_{i,2j} & \cdots & \Delta X_{n,2j} \\ \vdots & & \vdots & & \vdots \\ \Delta X_{1,2m} & \cdots & \Delta X_{i,2m} & \cdots & \Delta X_{n,2m} \end{bmatrix} \tag{13}$$

where $X_{i,j}$ and $\Delta X_{i,j}$ define the $j^{\text{th}}$ component of the Fourier coefficients for the $i^{\text{th}}$ degree-of-freedom for the non-linear system. The number of harmonic coefficients $m$ is selected in order to consider only the significant harmonics expected in the solution. By replacing $\mathbf{x}$ and $\Delta\mathbf{x}$ by their Fourier series, one obtains $(2m+1)\times n$ linear algebraic equations

$$\mathbf{AX} + \mathbf{F} - \mathbf{G} + \mathbf{A}.\Delta\mathbf{X} = \mathbf{0} \tag{14}$$

with

$$\mathbf{A} = \begin{bmatrix} \mathbf{K} & & & & \\ & \begin{bmatrix} -\Omega^2\mathbf{M}+\mathbf{K} & \Omega\mathbf{D} \\ -\Omega\mathbf{D} & -\Omega^2\mathbf{M}+\mathbf{K} \end{bmatrix} & & & \\ & & \ddots & & \\ & & & \begin{bmatrix} -j^2\Omega^2\mathbf{M}+\mathbf{K} & j\Omega\mathbf{D} \\ -j\Omega\mathbf{D} & -j^2\Omega^2\mathbf{M}+\mathbf{K} \end{bmatrix} & \\ & & & & \ddots \\ & & & & & \begin{bmatrix} -m^2\Omega^2\mathbf{M}+\mathbf{K} & m\Omega\mathbf{D} \\ -m\Omega\mathbf{D} & -m^2\Omega^2\mathbf{M}+\mathbf{K} \end{bmatrix} \end{bmatrix} \tag{15}$$

$\mathbf{F}$ and $\mathbf{G}$ represent the Fourier coefficients of the function $\mathbf{f}$ and $\mathbf{g}$, respectively. $\mathbf{F}$ is difficult to determine from the Fourier coefficients directly due to the dependence in time for this function. Hence $\mathbf{F}$ can be calculated by using the following path (Alternating Frequency/Time domain AFT) as follows

$$\mathbf{X} \xrightarrow{\Gamma^{-1}} \mathbf{x}(t) \longrightarrow \mathbf{f}(t) \xrightarrow{\Gamma} \mathbf{F} \tag{16}$$

where $\Gamma$ and $\Gamma^{-1}$ define the Discrete Fourier Transform from time to frequency domain, and from frequency to time domain, respectively. The DFT from time to frequency domain is given by

$$\Gamma_{ij} = \begin{cases} \dfrac{1}{2m+1} & \text{for } i=1 \\ \dfrac{2}{2m+1}\cos\left(\dfrac{(j-1)i\pi}{2m+1}\right) & \text{for } i=2,4,\ldots,2m \\ \dfrac{2}{2m+1}\sin\left(\dfrac{(j-1)(i-1)\pi}{2m+1}\right) & \text{for } i=3,\ldots,2m+1 \end{cases} \quad \text{for } j=1,2,\ldots,2m+1 \tag{17}$$

and from frequency time domain

$$\Gamma_{ij}^{-1} = \begin{cases} 1 & \text{for} \quad j = 1 \\ \cos\left(\dfrac{(i-1)\,j\pi}{2m+1}\right) & \text{for} \quad j = 2,4,...,2m \\ \sin\left(\dfrac{(i-1)(j-1)\pi}{2m+1}\right) & \text{for} \quad j = 3,...,2m+1 \end{cases} \qquad \text{for } i = 1,2,...,2m+1 . \tag{18}$$

Finally, the error vector $\mathbf{R}$ is given by

$$\mathbf{R} = \mathbf{AX} + \mathbf{F} - \mathbf{G} \tag{19}$$

and the associated convergence values are defined by

$$\delta_1 = \sqrt{\mathbf{R}_0^2 + \sum_{j=1}^{m}\left(\mathbf{R}_{2j-1}^2 + \mathbf{R}_{2j}^2\right)} \tag{20}$$

and

$$\delta_2 = \sqrt{\Delta\mathbf{X}_0^2 + \sum_{j=1}^{m}\left(\Delta\mathbf{X}_{2j-1}^2 + \Delta\mathbf{X}_{2j}^2\right)} \tag{21}$$

The global procedure of the Alternate/Frequency Time domain method is illustrated in Figure 4.

## 4.2. PATH CONTINUATION

Usually, the dynamics of system and the solution associated have to be calculated at different parameter values consecutively (in this study, the considered parameter is the speed of shaft rotation). In order to reduce the time required for the calculation the path following technique can be used (Narayanan and Sekar [38]). One considers the estimation of the neighboring point on the solution branch by using the Lagrangian polynomial extrapolation method with four points. So, one assumes that four points on the solution branch are obtained a priori in order to begin the extrapolation scheme. The varied parameter is the speed rotation of the shaft $\Omega$. Any point on the solution branch is represented at $\mathbf{X}_i$, where $\mathbf{X}_i$ is the Fourier coefficients of $\mathbf{x}$. The arc length between two consecutive points $\mathbf{X}_{i+1}$ and $\mathbf{X}_i$ is given by $\delta l_{i+1} = \sqrt{\left(\mathbf{X}_{i+1} - \mathbf{X}_i\right)^T\left(\mathbf{X}_{i+1} - \mathbf{X}_i\right)}$ for $i = 0,\ldots,2$. Next, the arc length parameters are calculated as follows

$$\begin{cases} L_0 = 0 \\ L_1 = \delta l_1 \\ L_2 = L_1 + \delta l_2 \\ L_3 = L_2 + \delta l_3 \\ L_4 = L_3 + \Delta l \end{cases} \tag{22}$$

Finally, by using the Lagrangian extrapolation scheme, the following estimated point at the distance $\Delta l$ might be defined by

$$\mathbf{X}_4 = \sum_{i=1}^{3}\left(\prod_{\substack{j=0 \\ i\neq j}}^{3}\left(\frac{L_3 - L_j}{L_i - L_j}\right)\mathbf{X}_i\right) \qquad \text{for } i = 0,\ldots,3 . \tag{23}$$

4.3. RESULTS AND DISCUSSIONS

The dynamic response of the rotor with the breathing crack is obtained by using the AFT method and the path following procedure. The Fourier series is truncated in order to consider only the significant harmonics. All the values of the parameters (damping, non-dimensional crack depth $\mu$, mass unbalance, eccentricity of the mass unbalance, etc…) are given in Table 1. It may be noted that the crack depth, the level of damping, and the level of unbalance are important parameters because these three all influence the magnitude of the non-linear term

In order to demonstrate the capability of the Alternate/Frequency Time domain method for determining a good approximation of the dynamic response of the rotor with a breathing crack, a comparison is performed with results obtained by using direct numerical integration such as the Runge-Kutta 4. Figure 5 shows the comparison between times histories of the vertical and horizontal displacements of the cracked rotor around the crack position ($L = 0.375\mathrm{m}$), evaluated through numerical integration and the AFT method for various numbers of harmonics $m$. The associated orbits are plotted in Figure 6. These results highlight that the periodic solution ($m = 1$) does not provide a good approximation of the solution but that the second harmonic coefficients ($m = 2$) are sufficient to obtain an accurate evaluation of the dynamic response of the rotor with a breathing crack. Therefore, these results clearly demonstrate that the presence of a breathing crack results in a non-linear dynamical behavior.

Figure 7 shows the horizontal and vertical steady-state responses of the cracked and uncracked shafts, at the position of the shaft $L = 0.375\mathrm{m}$. Moreover Figure 8 shows the horizontal and vertical steady-state responses of the cracked shaft for each node of the shaft. Figure 7 indicates a decrease in the first and second vertical natural frequencies as indicated at the marks 3 and 5. This change is due to the reduction in system stiffness resulting from the presence of the transverse crack. Although the first and second horizontal natural frequencies decrease due to the presence of the crack, it may be observed that the differences between the cracked and uncracked frequencies are very small and very difficult to detect in practice, as indicated at the marks 7 and 8 in the Figure 7.

In the case of the cracked shaft, 2X harmonics are observed in the horizontal and vertical directions at one-half of the first vertical and horizontal frequencies, as illustrated at the marks 2 and 6 in Figures 7 and 8. 3X harmonics in the vertical direction for the first frequency are also present at one-third of the first vertical natural frequency as shown in Figures 7 and 8 at the mark 1. Moreover, 2X harmonics in the vertical direction for the second vertical frequency at one-half of the second vertical frequency. It should be noted that there are no visible 3X harmonics of the first horizontal natural frequency and 2X harmonic of the second horizontal natural frequency. Considering the rotor without a crack, no harmonics are predicted for the dynamic behavior of the rotor. In this case, the dynamic response is purely synchronous.

In order to compare the dynamic behavior of the cracked and uncracked shaft near one-half and one-third of the natural frequencies, the differences between the cracked and uncracked horizontal and vertical amplitudes are calculated as follows

$$\Delta X = \left|\mathrm{Max}\left(X_{cracked} - X_{uncracked}\right)\right| \tag{24}$$

and

$$\Delta Y = \left|\mathrm{Max}\left(Y_{cracked} - Y_{uncracked}\right)\right| \tag{25}$$

at each element of the shaft. $X_{cracked}$, $Y_{cracked}$, $X_{uncracked}$, and $Y_{uncracked}$ define the horizontal and vertical displacement for the cracked and uncracked shaft respectively. Therefore the Figure 9 shows clearly the differences $\Delta X$ and $\Delta Y$ between the cracked and uncracked rotor. As illustrated in Figure 9(A) and 9(B), the differences $\Delta X$ and $\Delta Y$ increase at the one-half and one-third of the first horizontal and vertical frequencies. Moreover, the differences $\Delta X$ and $\Delta Y$ at each element of the shaft indicate clearly the first mode shape with the maximum difference obtained at the middle of the shaft. Figure 9(C) shows the differences $\Delta X$ and $\Delta Y$ around one-half of the second horizontal and vertical frequencies. In this case the differences $\Delta X$ and $\Delta Y$ also increase one-half of the second horizontal and vertical frequencies and indicate the deformations of the second mode shape with a node in the middle of the shaft and the maximums of $\Delta X$ and $\Delta Y$ are obtained at one fourth on each end of the shaft.

As explained previously, the diagnosis of the presence of a crack in rotating machinery based only on the appearance of a harmonics response at half the natural frequency in the spectrum may be misleading. So one of the features to detect the presence of a transversal crack in a rotating shaft is the use of the evolution of the orbits during time around one-half of the resonance frequencies.

Now, the attention is focussed on the evolution of the orbits around one-half of the natural frequencies. Figure 10 shows the evolution of the rotor's orbits. It is shown that a distortion in the orbit appears and increases when the speed of the rotor increases. Next, the shape of the orbit changes from a form with only one loop to a double loop. And finally, when the speed of the cracked rotor is passing through nearly half of the critical speed, the orbit changes from a double loop to a inside loop (a loop containing another small loop inside) that indicates the change in both the phase and the amplitude of the harmonic components. This dynamic behavior of the rotor through the passage of half of the rotor speeds is the signature of a crack in the rotor. Moreover, Figure 11 shows the orbit obtained close to a third of the first critical speed, half of the first critical speed, and half of the second critical speed. It may be observed that the change in the dynamic behavior is also detected near half of the second critical speed with the formation of double loops at one end of the shaft and a distortion form on the other end (Figure 11(c)). Showing the dynamic behavior at third of the first critical speed (Figure 11(a)), the change of the orbit appears clearly with a triple loops that indicates the presence of the harmonics. All these changes of the dynamic behavior of the rotor near half and third of critical speeds may be considered as a feature for detection of crack in the rotor shafts.

Finally a parametric study is presented with various crack depth are performed in order to show the influence of the crack depth by considering the variation and change of orbit around one-half of the first frequency. As illustrated in Figure 12, the orbits decrease when the non-dimensional crack depth $\mu$ decreases. Moreover, the frequency at which 2X harmonic occurs increases as the crack depth decreases, due to the reduced stiffness resulting from the crack, which changes the natural frequencies. This is illustrated by plotting (in Figure 13) the differences $\Delta X$ and $\Delta Y$ at the middle position of the shaft for the various crack depth: as it can be seen the position of the maximum differences versus the frequency increase with the decreasing of the crack depth, as well as the differences $\Delta X$ and $\Delta Y$ decrease. Therefore, the behavior of the cracked rotor around one-half of the natural frequencies and the observation of the associated orbits could provide crack diagnostic information and could permit the estimation of the crack depth.

## 5. CONCLUSION

In this study, the influence of transversal cracks has been investigated: the change of the shaft frequencies, as well as the harmonic component of the dynamical system response and the evolution of the orbits are the principal effects due to the presence of a crack in a rotating shaft. More particularly, the changes in the non-linear dynamical behavior of the rotor system through half resonance speeds appear to be the classical signature for detecting the presence of a breathing crack. Indeed, the distortion of the orbit, and formation of a double loop and inside loop in the orbit could be considered as one of the most practical indicators of the presence of a transversal crack for health monitoring purposes. Moreover the observation of the orbit amplitudes at half resonance speeds could also provide crack diagnostic information about the crack depth: when the crack depth increases, an increase of the orbit amplitudes is observed, as well as a decrease of the shaft speed at which the 2X harmonic component of the dynamic response is maximum

In this paper, the use of the Alternate Frequency/Time Domain method and a path following procedure allows obtaining rapidly and efficiently the non-linear dynamical behavior of a rotating shaft with a transverse breathing crack. This method has advantages in terms of computating time. It is easily implemented.

## APPENDIX A: MOMENTS ABOUT THE CENTROID FOR THE CRACKED ROTOR

As illustrated in Figure 1, the cross section of the shaft at the location of the crack has asymmetric area moments inertia about the neutral axis of bending. The area moments of inertia $\tilde{I}_X$ and $\tilde{I}_Y$ about the *X* and *Y*-axes are defined as

$$\tilde{I}_X = \int_A Y^2 dA = \int_A Y^2 dXdY \tag{26}$$

$$\tilde{I}_Y = \int_A X^2 dA = \int_A X^2 dXdY \tag{27}$$

where *A* defines the uncracked area of the cross section. After integrating over the uncracked area, one obtains

$$\tilde{I}_X = \frac{R^4}{4}\left[(1-\mu)\left(1-4\mu+2\mu^2\right)\sqrt{2\mu - \mu^2} + \cos^{-1}(1-\mu)\right] \tag{28}$$

$$\tilde{I}_Y = \frac{\pi R^4}{4} + R^4\left[\frac{2}{3}(1-\mu)\left(2\mu - \mu^2\right)^{3/2} + \frac{1}{4}(1-\mu)\left(1-4\mu+2\mu^2\right)\sqrt{\left(2\mu - \mu^2\right)} + \sin^{-1}\left(\sqrt{\left(2\mu - \mu^2\right)}\right)\right] \tag{29}$$

where $R$ and $\mu$ are the shaft radius and the nondimensional crack depth ($\mu = a/R$). Then, the moment of inertia about the centroidal axes $I_X$ and $I_Y$ are obtained

$$I_X = \tilde{I}_X \tag{30}$$

$$I_Y = \tilde{I}_Y - A\bar{X}^2 \tag{31}$$

where $\bar{X}$ defines the centroid of the cross section. The uncracked area of the cross section *A*, and the distance from the axis $X$ to the centroid of the cross section $\bar{X}$ are given by

$$A = R^2\left[(1-\mu)\sqrt{2\mu - \mu^2} + \cos^{-1}(1-\mu)\right] \tag{32}$$

$$\bar{X} = \frac{2}{3A}R^3\left(2\mu - \mu^2\right)^{3/2}. \tag{33}$$

## APPENDIX B: NOMENCLATURE

| | |
|---|---|
| $\mu$ | non-dimensional crack depth |
| $\Omega$ | rotational speed |
| $e$ | eccentricity of the unbalance mass |
| $L_{\text{crack}}$ | position of the crack |
| $f_i^{\text{cracked}}$ | $i^{\text{th}}$ natural frequency for the cracked system |
| $f_i^{\text{uncracked}}$ | $i^{\text{th}}$ natural frequency for the uncracked system |
| $m$ | number of Fourier coefficients retained |
| $\%C_i$ | percentage change between the $i^{\text{th}}$ cracked and uncracked natural frequencies |
| $\mathbf{x}$ | displacement of the degree-of-freedom |
| $\dot{\mathbf{x}}$ | velocity of the degree-of-freedom |
| $\ddot{\mathbf{x}}$ | acceleration of the degree-of-freedom |
| **M** | mass matrix of the rotor system |
| **K** | stiffness matrix of the rotor system |
| $\mathbf{K}_{crack}$ | stiffness matrix for the cracked element |
| **C** | damping matrix of the rotor system |
| **G** | gyroscopic matrix of the rotor system |
| **D** | global damping matrix of the rotor system |
| **Q** | vector of gravity force for the rotor system |
| **W** | vector of imbalance force for the rotor system |
| **X** | matrix of the Fourier coefficients associated with the displacements of the rotor system |

| Parameters | Physical dimension |
|---|---|
| Length of the shaft | 1 m |
| Diameter of the shaft | 0.01 m |
| Young's modulus of elasticity $E$ | $2.1\times10^{11}$ N.m$^{-2}$ |
| Shear modulus $G$ | $7.7\times10^{10}$ N.m$^{-2}$ |
| Poisson ratio $\nu$ | 0.3 |
| Density $\rho$ | 7800 kg.m$^{-3}$ |
| Position of disc 1 | 0.3 m |
| 5.1. POSITION OF DISC 2 | 0.7 m |
| Outer diameter of discs 1 and 2 | 0.04 m |
| Inner diameter of discs 1 and 2 | 0.01 m |
| Thickness of discs 1 and 2 | 0.015 m |
| Location of the crack $L_{crack}$ | 0.375 m |
| Non-dimensional crack depth $\mu$ | 1 |
| Mass unbalance | 0.005 g |
| Phase unbalance | 0 degree |
| Eccentricity of the mass unbalance | 0.02 m |
| Coefficient of damping $\beta$ | $10^{-5}$ |

Table 1: Details of the rotor model

| | uncracked | $\mu = 0.25$ | $\mu = 0.5$ | $\mu = 0.75$ | $\mu = 1$ |
|---|---|---|---|---|---|
| 1st frequency (Hz) | 16.597 | 16.581 | 16.549 | 16.479 | 16.257 |
| 2nd frequency (Hz) | 16.597 | 16.582 | 16.559 | 16.543 | 16.541 |
| 3rd frequency (Hz) | 65.367 | 65.290 | 65.131 | 64.787 | 63.755 |
| 4th frequency (Hz) | 65.367 | 65.292 | 65.178 | 65.102 | 65.088 |
| 5th frequency (Hz) | 176.038 | 176.032 | 176.015 | 175.980 | 175.866 |
| 6th frequency (Hz) | 176.038 | 176.032 | 176.020 | 176.012 | 176.011 |

Table 2: Evolution of the natural frequencies with the non-dimensional crack depth ( $L_{crack} = 0.375\text{m}$ )

| | uncracked | Position of the crack $L_{crack}$ | | | | |
|---|---|---|---|---|---|---|
| | | 0.025m | 0.225m | 0.475m | 0.725m | 0.975m |
| 1st frequency (Hz) | 16.597 | 16.593 | 16.399 | 16.169 | 16.332 | 16.593 |
| 2nd frequency (Hz) | 16.597 | 16.597 | 16.564 | 16.525 | 16.553 | 16.597 |
| 3rd frequency (Hz) | 65.367 | 65.316 | 63.673 | 65.321 | 63.779 | 65.323 |
| 4th frequency (Hz) | 65.367 | 65.359 | 65.078 | 65.360 | 65.094 | 65.360 |
| 5th frequency (Hz) | 176.038 | 175.756 | 173.312 | 171.737 | 174.523 | 175.612 |
| 6th frequency (Hz) | 176.038 | 175.993 | 175.559 | 175.295 | 175.775 | 175.970 |

Table 3: Evolution of the natural frequencies with the location of the crack $L_{crack}$ ( non-dimensional crack depth $\mu = 1$ )

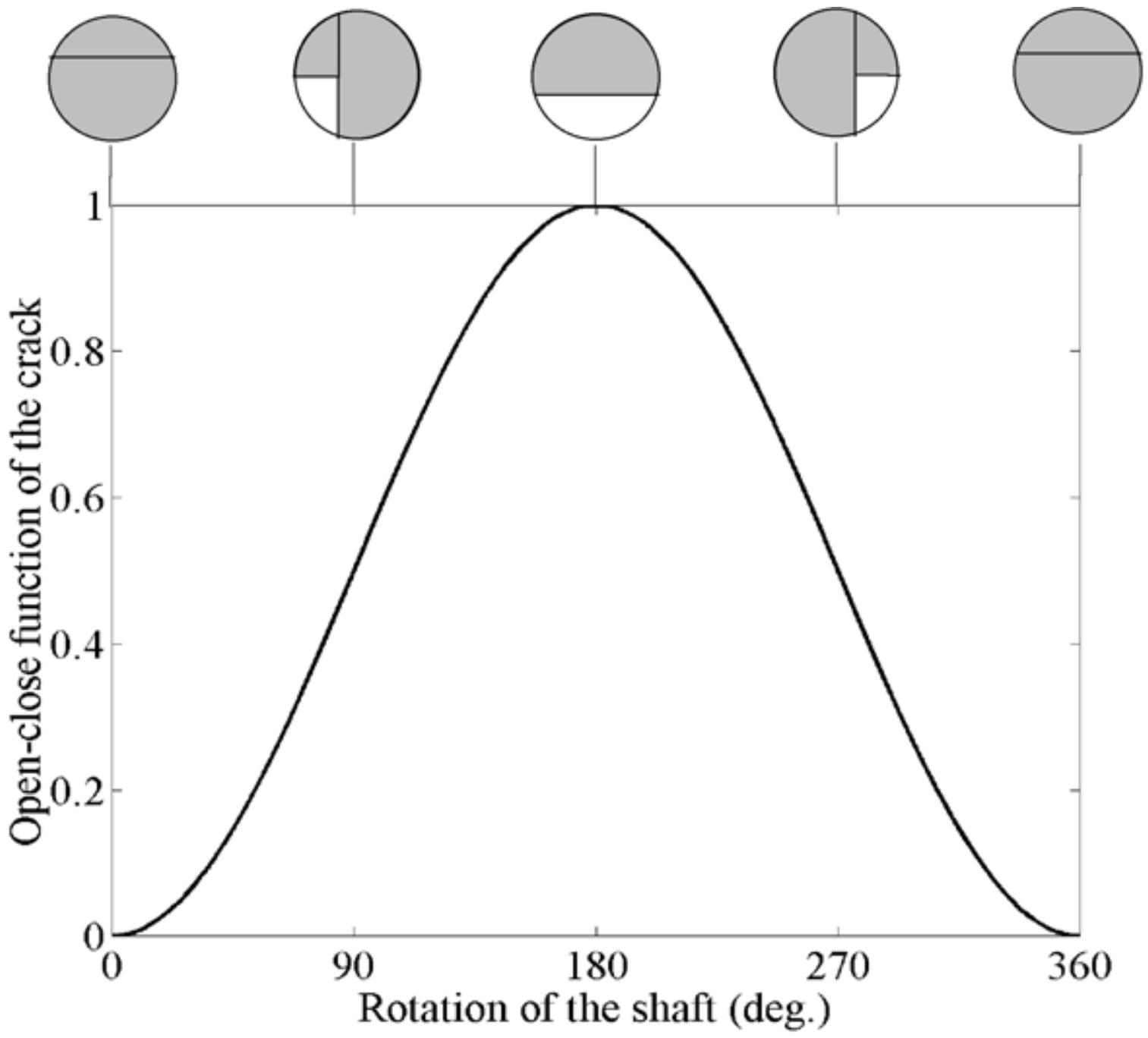

Figure 1 : Model of stiffness variation for the breathing crack

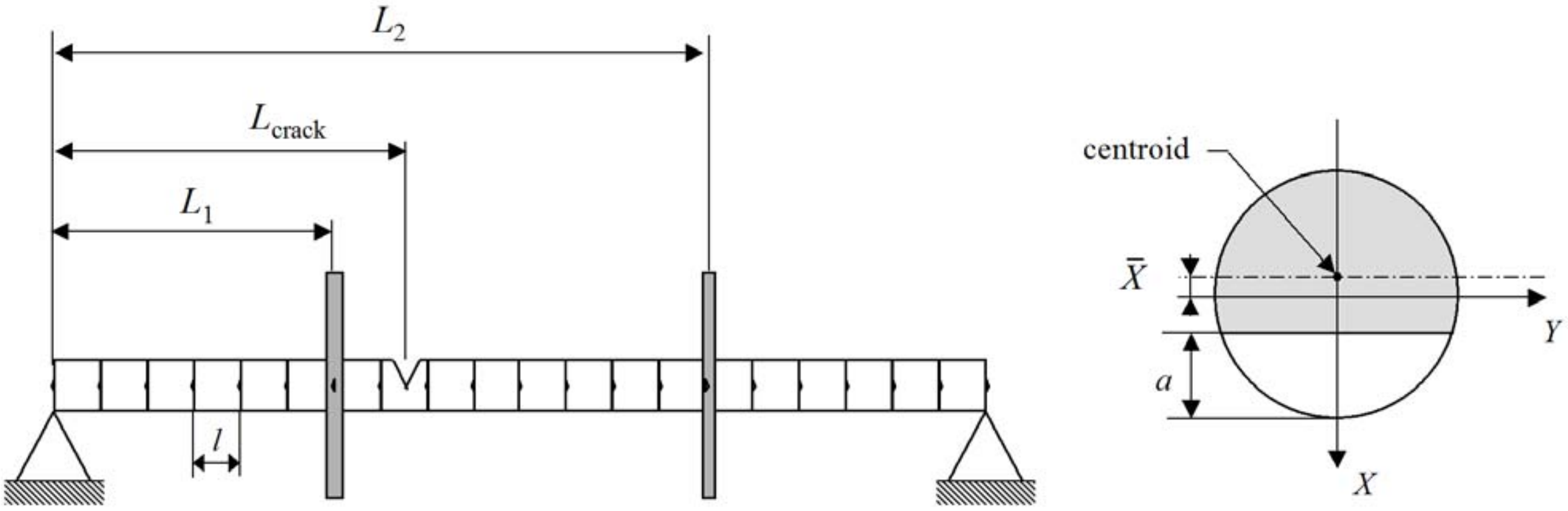

Figure 2 : Rotor model with two discs and a cracked element

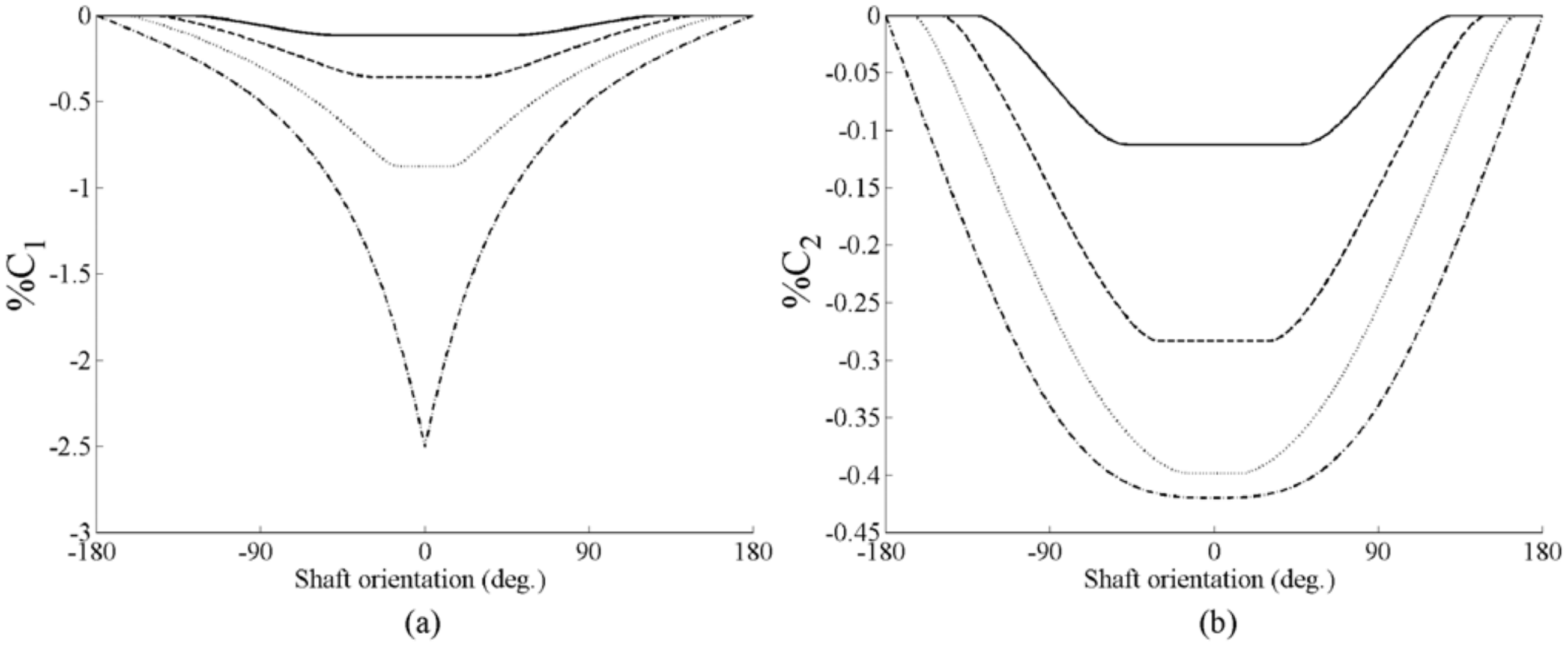

Figure 3 : Changes in the first and second natural frequencies with the rotation of the shaft and the non-dimensional crack depth $\mu$

( —— $\mu = 0.25$ , - - - - $\mu = 0.5$ , ·········· $\mu = 0.75$ , - · - · $\mu = 1$ )

(a) first frequency (b) second frequency

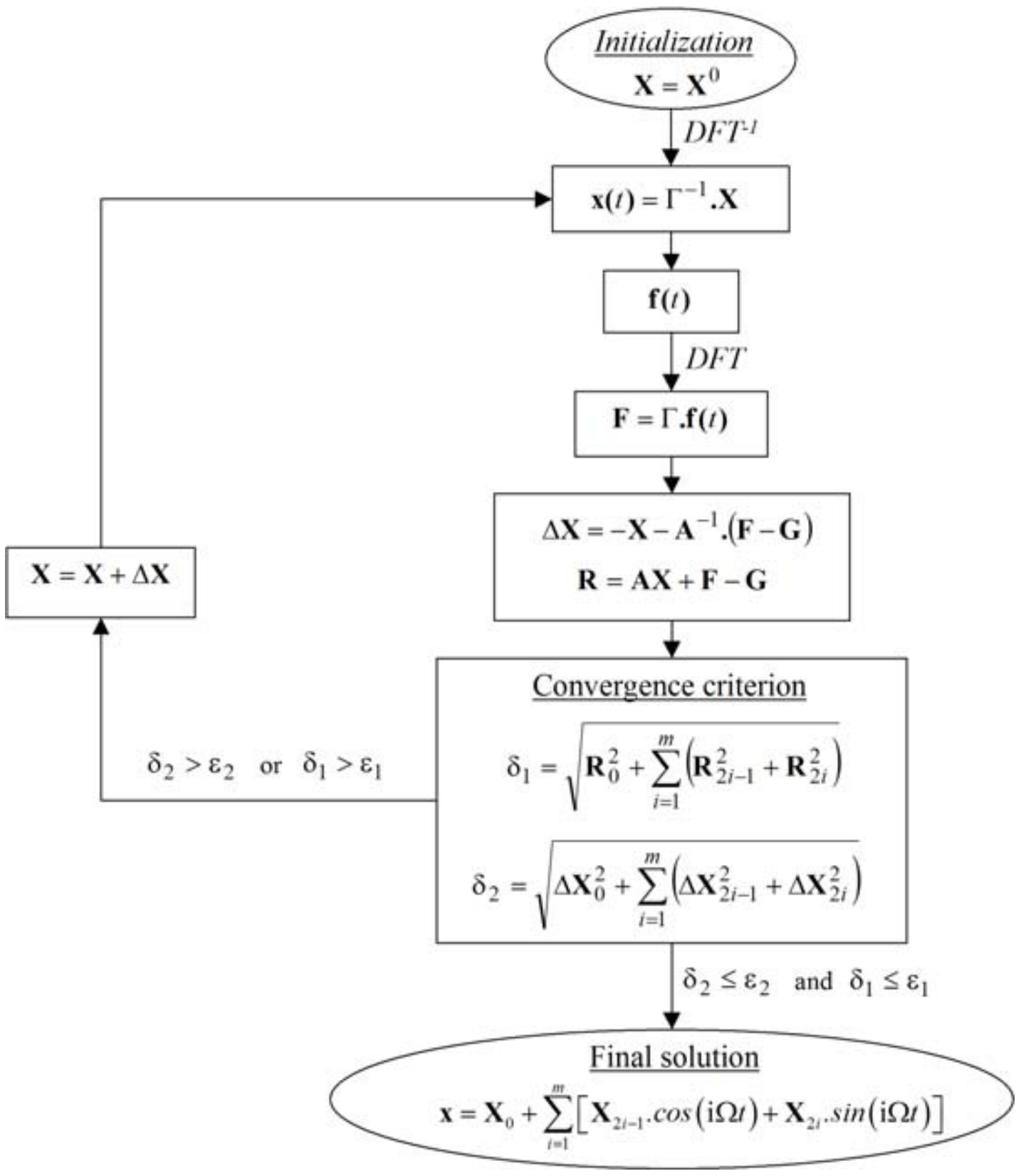

Figure 4 : Alternate Frequency/Time domain method

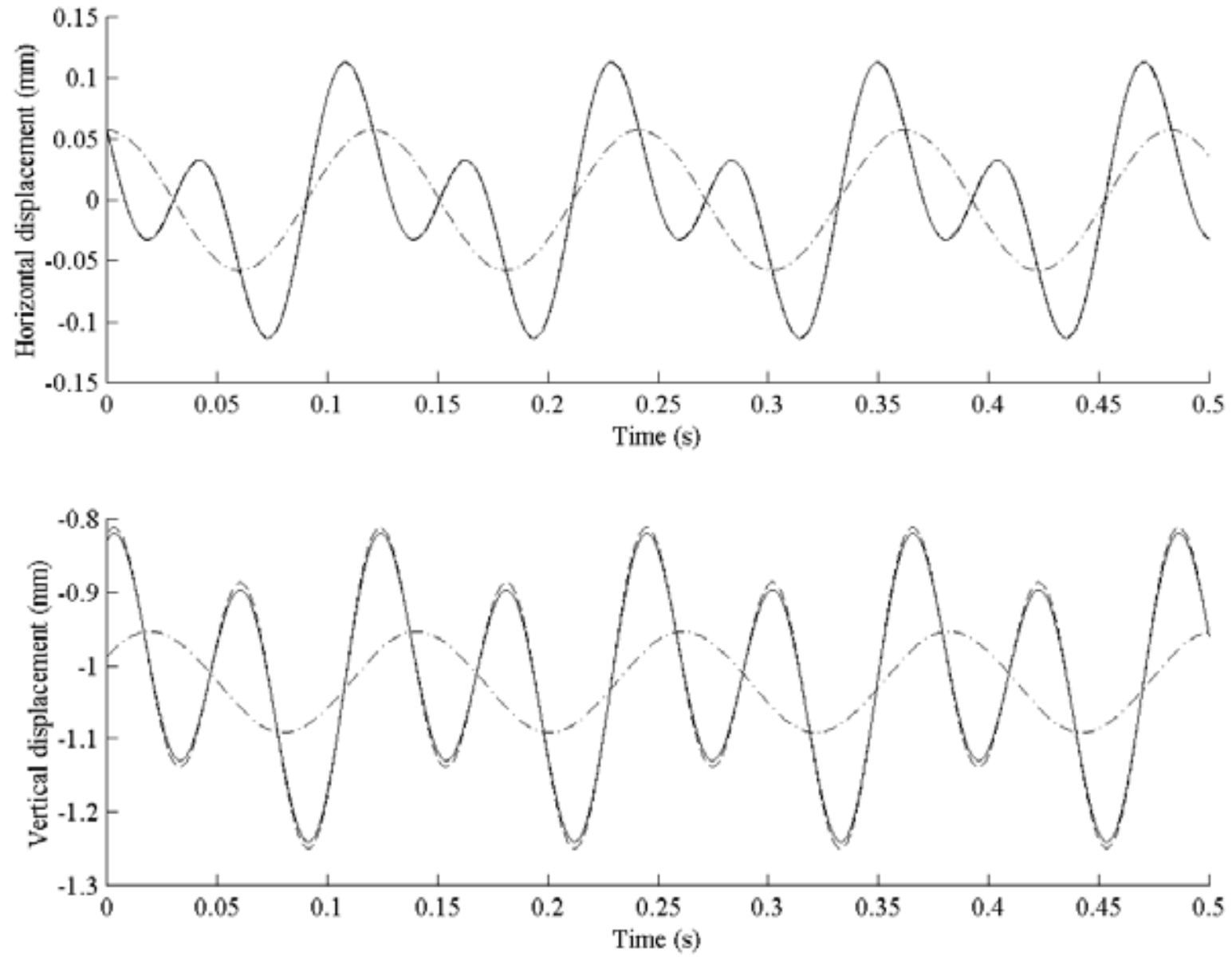

Figure 5 : Evolution of the horizontal and vertical displacements at the position of the crack ($L_{\text{crack}} = 0.35\text{m}$) for $f = 8.285\text{Hz}$
( —— Runge-Kutta 4, -·-·- *m*=1, - - - *m*=2)

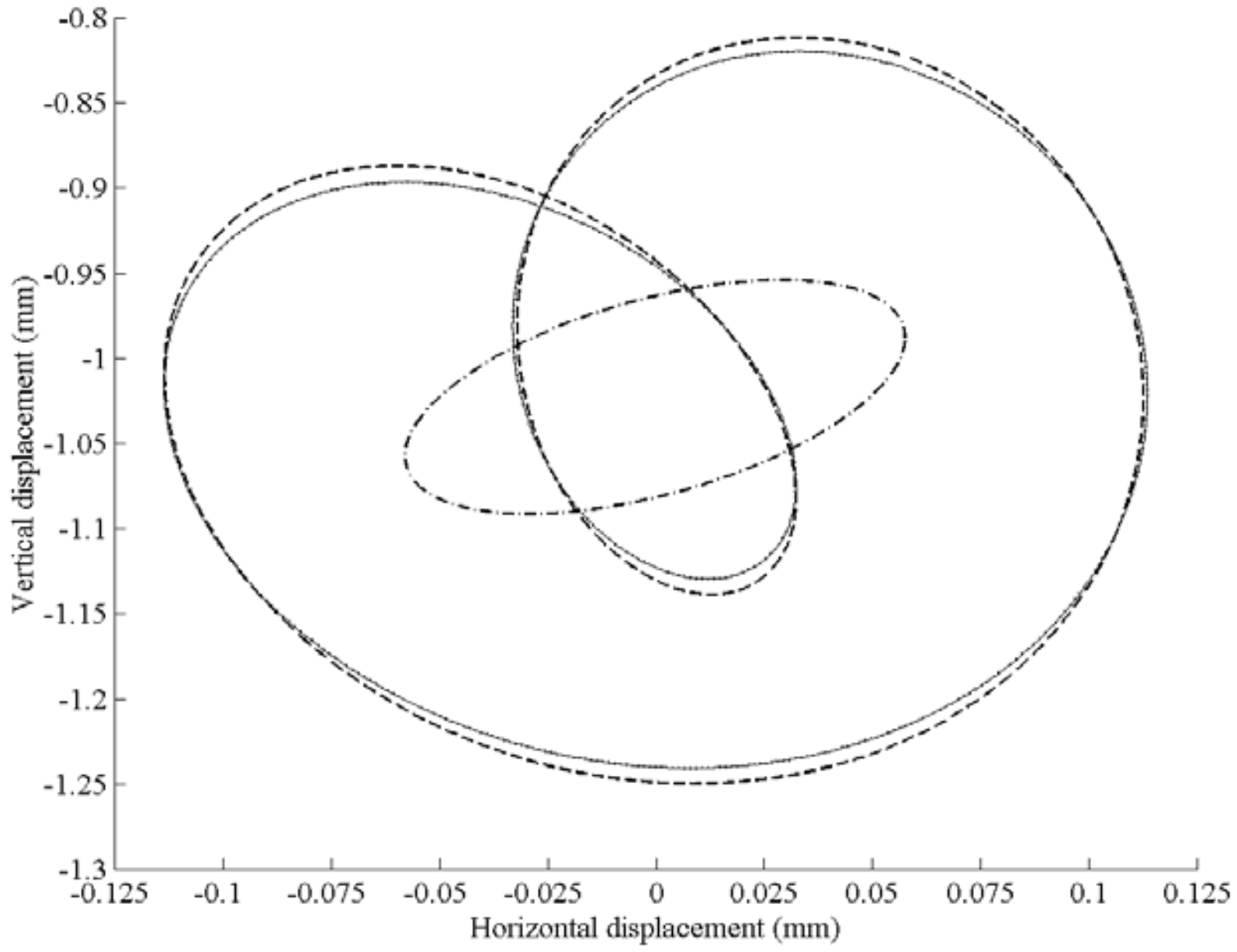

Figure 6 : Evolution of the orbit at the position of the crack ($L_{\text{crack}} = 0.35\text{m}$) for $f = 8.285\text{Hz}$
( —— Runge-Kutta 4, -·-·- *m*=1, - - - *m*=2)

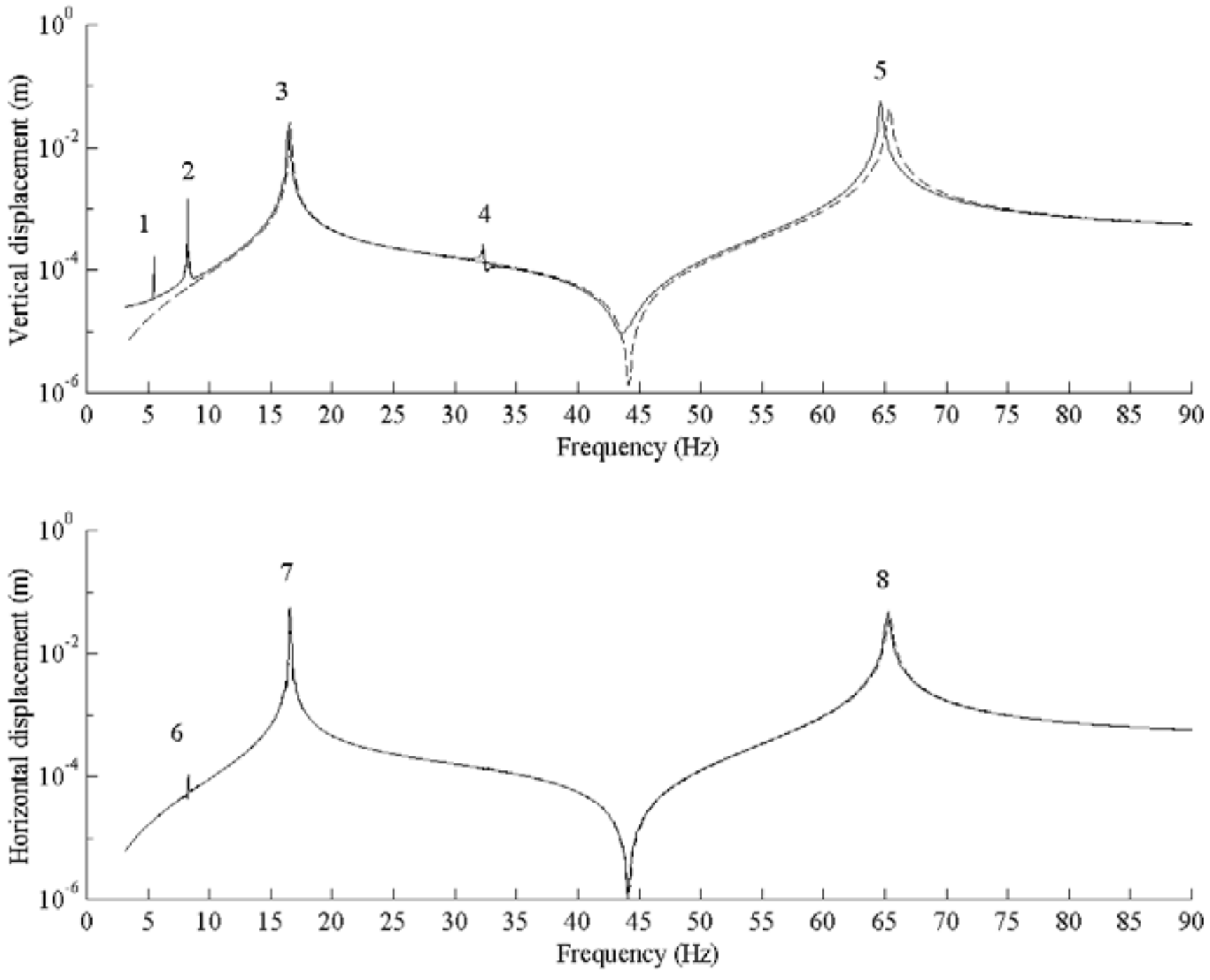

Figure 7 : Comparison between the cracked and uncracked shafts - horizontal and vertical displacements at the position $L = 0.35\text{m}$ ( - - - uncracked, —— cracked)

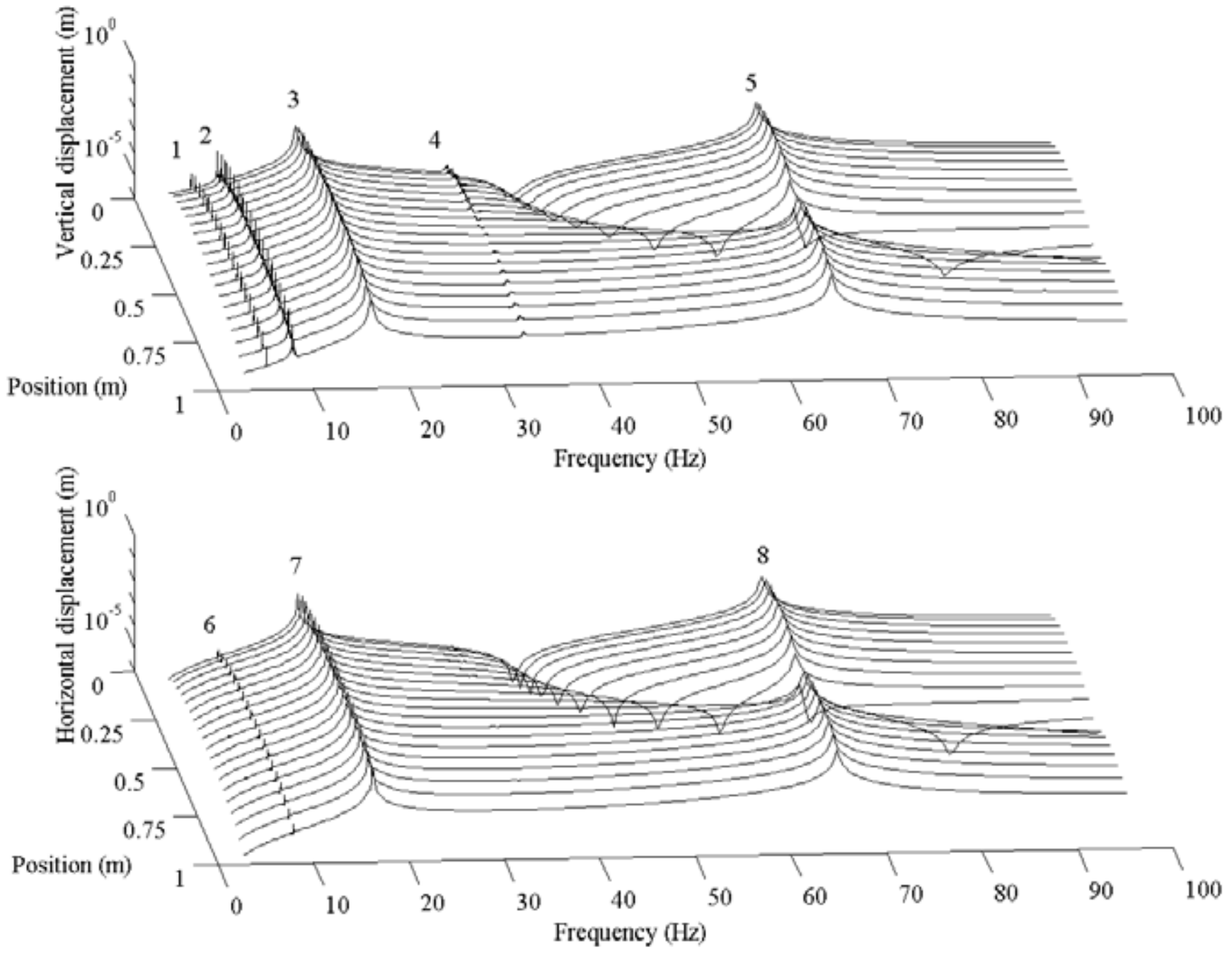

Figure 8 : Vertical and horizontal displacements for the cracked shaft

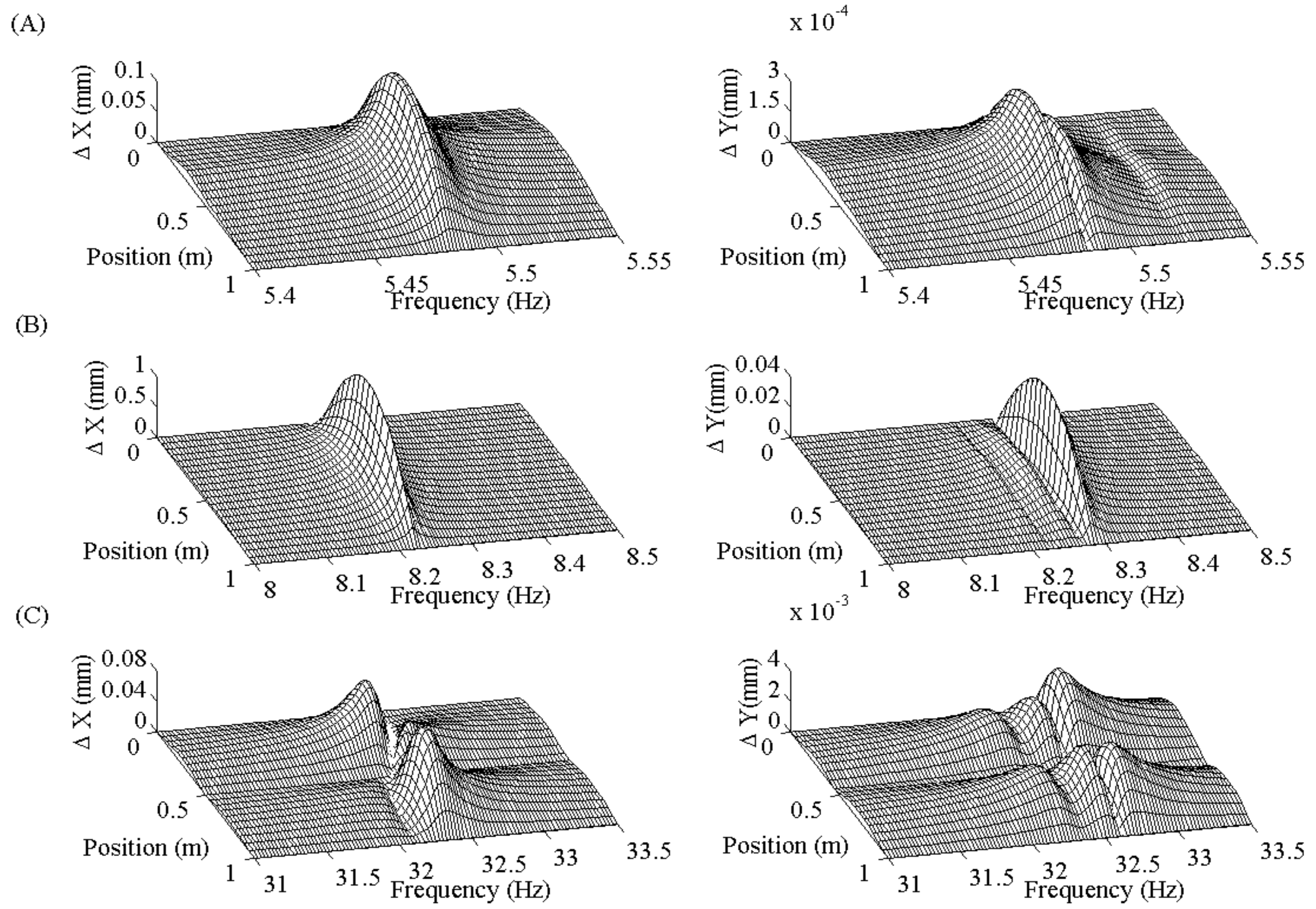

Figure 9 : Differences ΔX and ΔY between the cracked and uncracked shaft for (A) the 3X resonance of the first frequency (B) the 2X resonance of the first frequency and (C) the 2X resonance of the second frequency

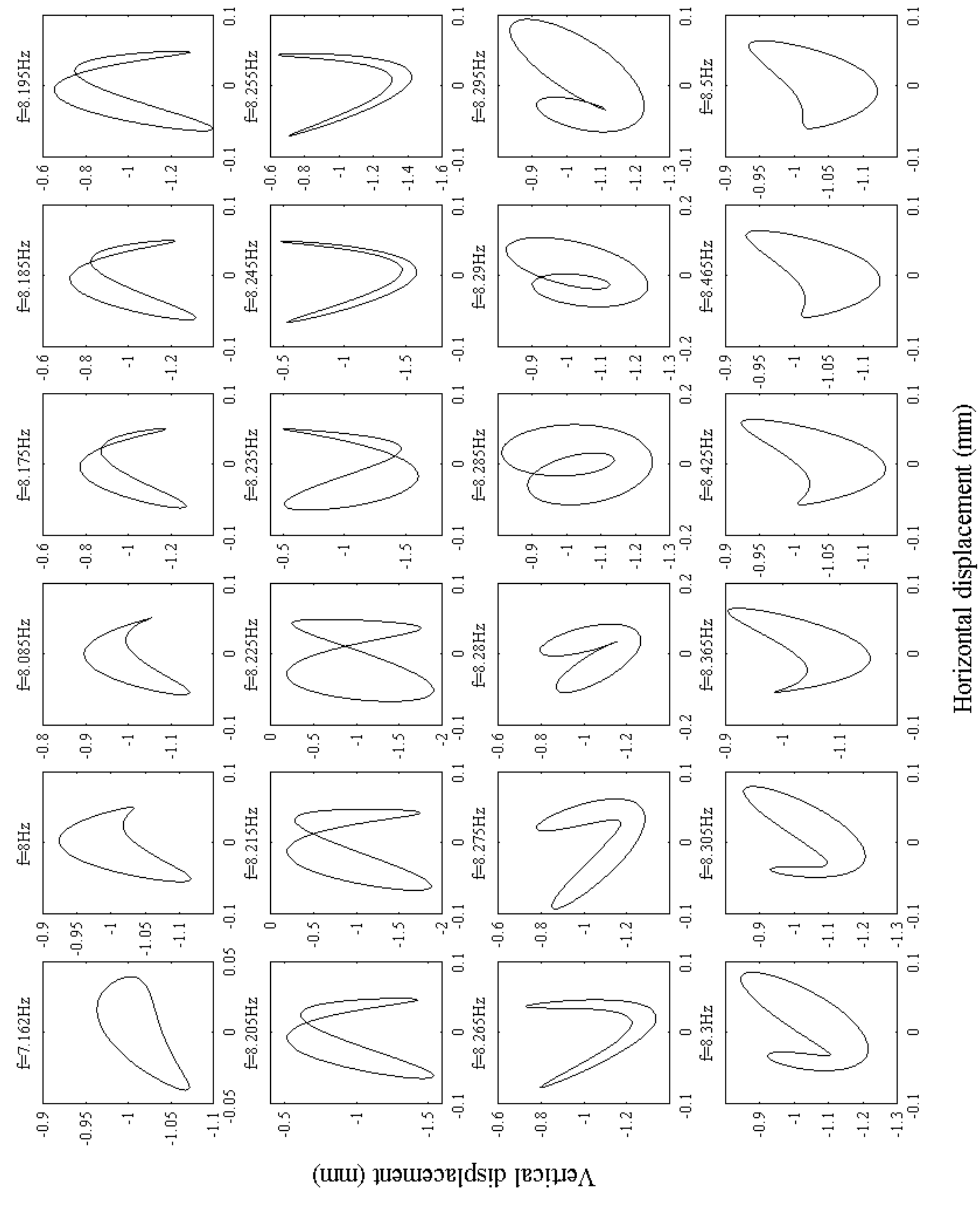

Figure 10 : Evolution of the orbit at the position of the crack ( $L_{crack} = 0.35\text{m}$ ) for the 2X resonance of the first frequency

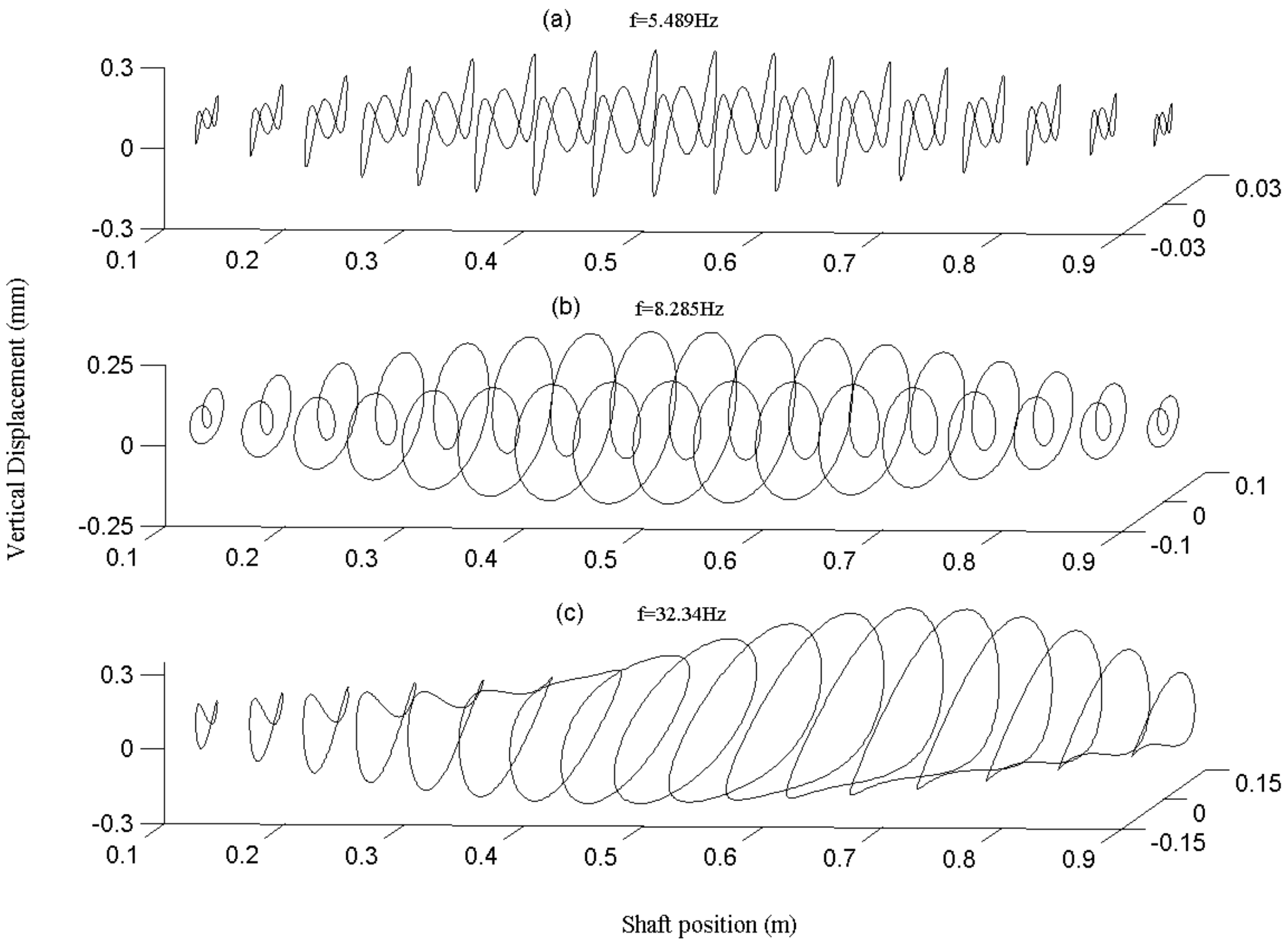

Figure 11 : Evolution of the orbits at various position on the shaft around the static deflection
(a) 3X resonance of the first frequency (b) 2X resonance of the first frequency
(c) 2X resonance of the third frequency

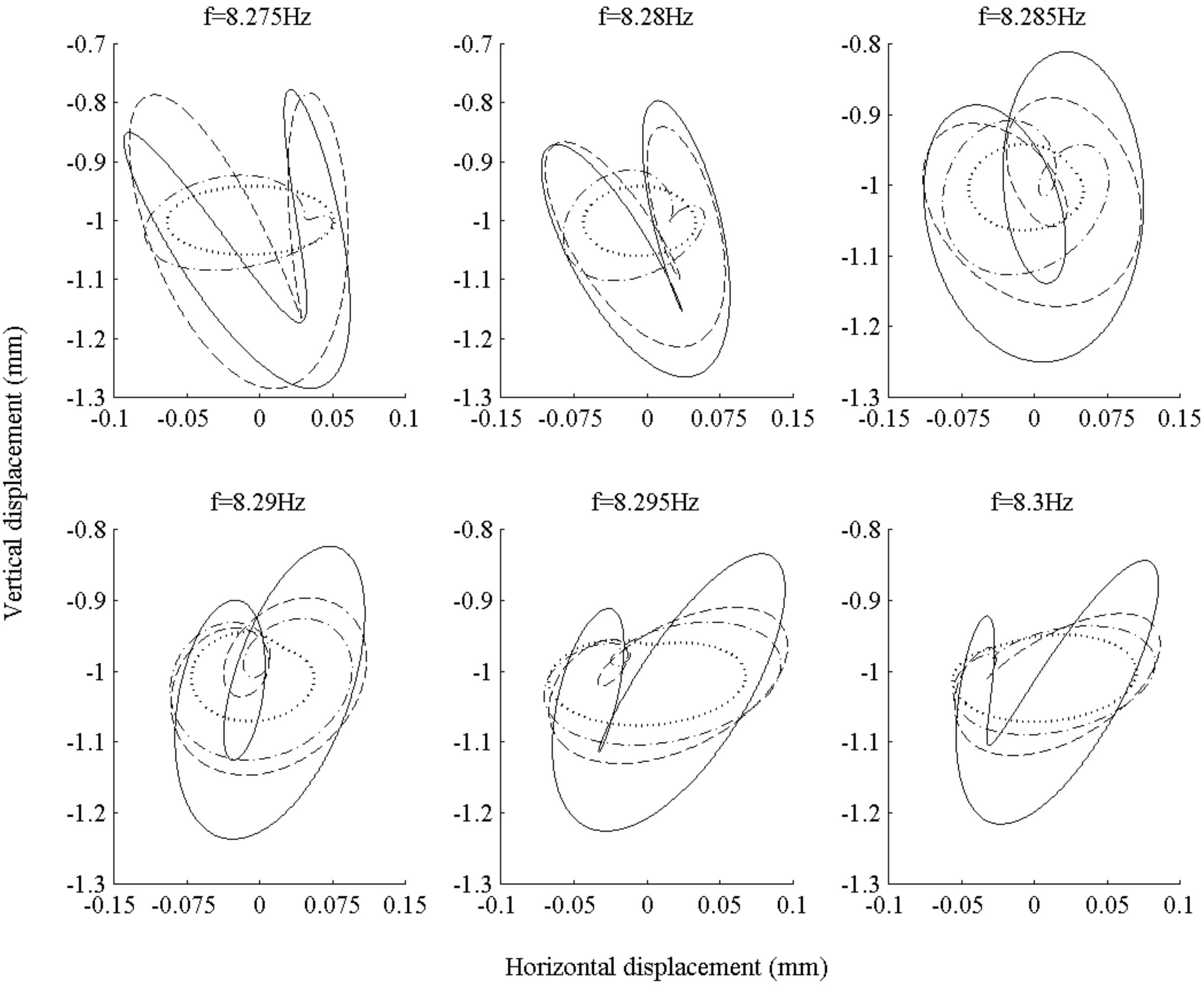

Figure 12 : Evolution of the orbit at the position of the crack ( $L_{\text{crack}} = 0.35\text{m}$ ) around the 2X resonance of the first frequency, with the variation of the non-dimensional crack depth μ

( ·········· $\mu = 0.25$ , ·−·− $\mu = 0.5$ , −−− $\mu = 0.75$ , ——— $\mu = 1$ )

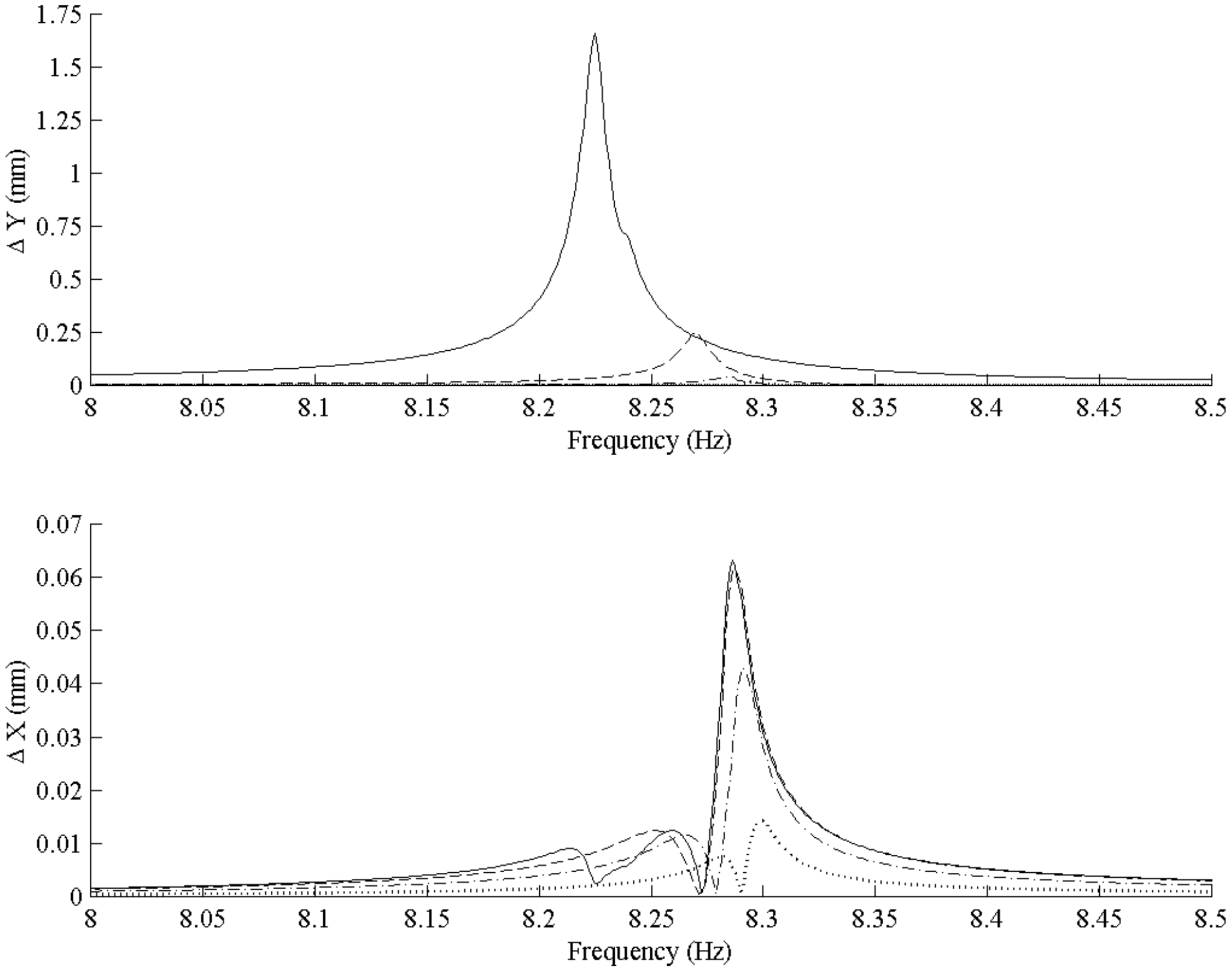

Figure 13 : Evolution of the differences $\Delta X$ and $\Delta Y$ between the cracked and uncracked shaft for the 2X resonance of the first frequency at the middle of the shaft with the variation of the non-dimensional crack depth $\mu$

( ·········· $\mu = 0.25$ , · – · – $\mu = 0.5$ , · – – – $\mu = 0.75$ , ——— $\mu = 1$ )